\def\3he{$^3$He}
\def\4he{$^4$He}
\def\6li{$^6$Li}
\def\7li{$^7$Li}
\def\la{\mathrel{\mathpalette\fun <}}

\def\fun#1#2{\lower3.6pt\vbox{\baselineskip0pt\lineskip.9pt
  \ialign{$\mathsurround=0pt#1\hfil##\hfil$\crcr#2\crcr\sim\crcr}}}
\def\beq#1\eeq{\begin{equation}#1\end{equation}}
\def\ie{{\it i.e.},}

\def\Yp{Y$_{\rm P}$}
\def\hi{H\thinspace{$\scriptstyle{\rm I}$}}
\def\hii{H\thinspace{$\scriptstyle{\rm II}$}}
\def\di{D\thinspace{$\scriptstyle{\rm I}$}}
\newcommand{\Deln}{\ensuremath{\Delta {\rm N}_\nu}}
\newcommand{\nnu}{\ensuremath{{\rm N}_\nu}}

\documentclass{iaus}
\usepackage{graphicx}

\title[IAUS 268.~~Primordial Nucleosynthesis] 
{Primordial Nucleosynthesis: \\ A Cosmological Probe}

\author[Gary Steigman]   
{Gary Steigman}

\affiliation{Departments of Physics and Astronomy \\ 
Center for Cosmology and Astro-Particle Physics \\
The Ohio State University \\
192 West Woodruff Avenue \\
Columbus, OH 43210 USA \\email: {\tt steigman@mps.ohio-state.edu}}

\pubyear{2010}
\volume{268}  
\jname{Light elements in the Universe}
\editors{C. Charbonnel, M. Tosi, F. Primas \& C. Chiappini, eds.}
\begin{document}

\maketitle

\begin{abstract}

During its early evolution the Universe provided a laboratory 
to probe fundamental physics at high energies.  Relics from 
those early epochs, such as the light elements synthesized 
during primordial nucleosynthesis when the Universe was only 
a few minutes old, and the cosmic background photons, last 
scattered when the protons (and alphas) and electrons 
(re)combined some 400 thousand years later, may be used to 
probe the standard models of cosmology and of particle physics.  
The internal consistency of primordial nucleosynthesis is tested 
by comparing the predicted and observed abundances of the light 
elements, and the consistency of the standard models is explored 
by comparing the values of the cosmological parameters inferred 
from primordial nucleosynthesis with those determined by studying 
the cosmic background radiation.  
\end{abstract}

\section{Introduction}

Primordial nucleosynthesis provides a key probe of the physics 
and early evolution of the Universe.   Big Bang Nucleosynthesis 
(BBN; $\sim 20$~minutes) and the cosmic microwave background 
(CMB) photons, last scattered at recombination ($\sim 400$~kyr), 
provide complementary probes of the physics of the early 
evolution of the Universe.

For a brief period during its early evolution the hot, dense 
Universe is a cosmic nuclear reactor.  Since the Universe 
is expanding and cooling rapidly, there is only time to 
synthesize in astrophysically interesting abundances the 
very lightest nuclides (D, \3he, \4he, and \7li).  In the 
standard models of cosmology and particle physics described 
by General Relativity, the universal expansion rate, the 
Hubble parameter, $H$, is determined by the total mass/energy 
density: $H^{2} \propto G\rho$, where $H = H(z)$, $z$ is the 
redshift, $G$ is the gravitational constant, and $\rho$ is 
the energy density.  During such early epochs, the Universe, 
which is filled with relativistic particles including three 
flavors of light neutrinos (\nnu~$= 3$), is ``radiation 
dominated", and the abundances of the nuclides synthesized 
during BBN depends on only one cosmological parameter, 
$\eta_{\rm B}$, which provides a measure of the universal 
density of baryons.
\beq
\eta_{\rm B} \equiv n_{\rm B}/n_{\gamma} \equiv 
10^{-10}\eta_{10}.
\label{eta}
\eeq
In eq.~\ref{eta}, $n_{\rm B}$ is the number density of baryons 
and $n_{\gamma}$ is the number density of cosmic background
photons.  The only baryons present at BBN are nucleons, 
\ie~protons and neutrons.  In contrast to the standard model 
of cosmology, there is a class of non-standard cosmological 
(and/or particle physics) models in which the expansion rate 
may differ from its standard model value, $H' \neq H$.  In these 
non-standard models the expansion rate can be parameterized 
by an ``expansion rate parameter", $S$, or equivalently, by 
an ``effective number of neutrinos", \nnu $\neq 3$, where
\beq
S^{2} \equiv (H'/H)^{2} \equiv G'\rho'/G\rho \equiv 1 + 
7\Delta{\rm N}_{\nu}/43.
\eeq
\label{nnu}
More generally, the effective number of ``extra" 
neutrinos, \Deln~$\equiv $~N$_{\nu} - 3$, parameterizes 
any non-standard energy density ($\rho' \neq \rho$), 
normalized to the contribution from one standard 
model neutrino by,
\beq
\Delta{\rm N}_{\nu} \equiv (\rho' - \rho)/\rho_{\nu}.
\eeq
\label{deltannu}
However, even if $\rho' = \rho$, it could be that 
\nnu~$\neq 3$ (\Deln~$\neq 0$) if the early-Universe 
gravitational constant differs from its current 
value, $G' \neq G$,
\beq
G'/G = S^{2} = 1 + 7\Delta{\rm N}_{\nu}/43.
\eeq
\label{G}
As will be seen below, in this class of non-standard 
models the BBN-predicted (nSBBN) primordial abundance of 
deuterium depends largely on the baryon density parameter, 
$\eta_{\rm B}$ (deuterium is a cosmological baryometer), 
while that of helium-4 is sensitive to the early Universe 
expansion rate, $S$ (\4he is an early universe chronometer).

In order to test the standard models of cosmology and 
particle physics, two key questions are addressed:\\
{\bf  1. Do the light element abundances predicted by 
BBN agree with the primordial abundances inferred 
from observations?}\\
{\bf  2. Do the BBN values of $\eta_{\rm B}$ and $S$ 
(\nnu) agree with those inferred from independent, 
non-BBN observations (e.g., from the CMB)?}

\section{Standard Big Bang Nucleosynthesis (SBBN)}
\label{sbbn}

For SBBN (\nnu~= 3), the light element relic abundances are 
only a function of the baryon density parameter, $\eta_{\rm B}$.
Among the light nuclides, deuterium is the baryometer of choice.
There are several reasons why D occupies this special place. 
One is that the post-BBN evolution of deuterium is simple and
monotonic: as gas is cycled through stars (producing the heavy
elements), D is only destroyed (\cite[Reeves \etal\ (1973)]{deut}, 
\cite[Epstein, Lattimer,  \& Schramm (1976)]{els}).  As a result, if 
deuterium is observed anywhere in the Universe, at any time in 
its evolution, the observed abundance will be no larger than the 
primordial value: (D/H)$_{\rm OBS} \leq $~(D/H)$_{\rm P}$.  In
addition, for systems of low metallicity, a sign that very little 
of their gas has been cycled through stars which destroy deuterium, 
the observed D abundance should approach the primordial value: 
(D/H)$_{\rm OBS} \rightarrow $~(D/H)$_{\rm P}$ (the ``Deuterium 
Plateau").  Another reason to prefer D is that its predicted 
primordial abundance is sensitive to the baryon density parameter; 
since (D/H)$_{\rm P} \propto \eta_{\rm B}^{-1.6}$, a $\sim 10\%$ 
determination of (D/H)$_{\rm P}$ results in a $\sim 6\%$ determination 
of $\eta_{\rm B}$.  

\begin{figure}[b]
\begin{center}
 \includegraphics[width=3.4in]{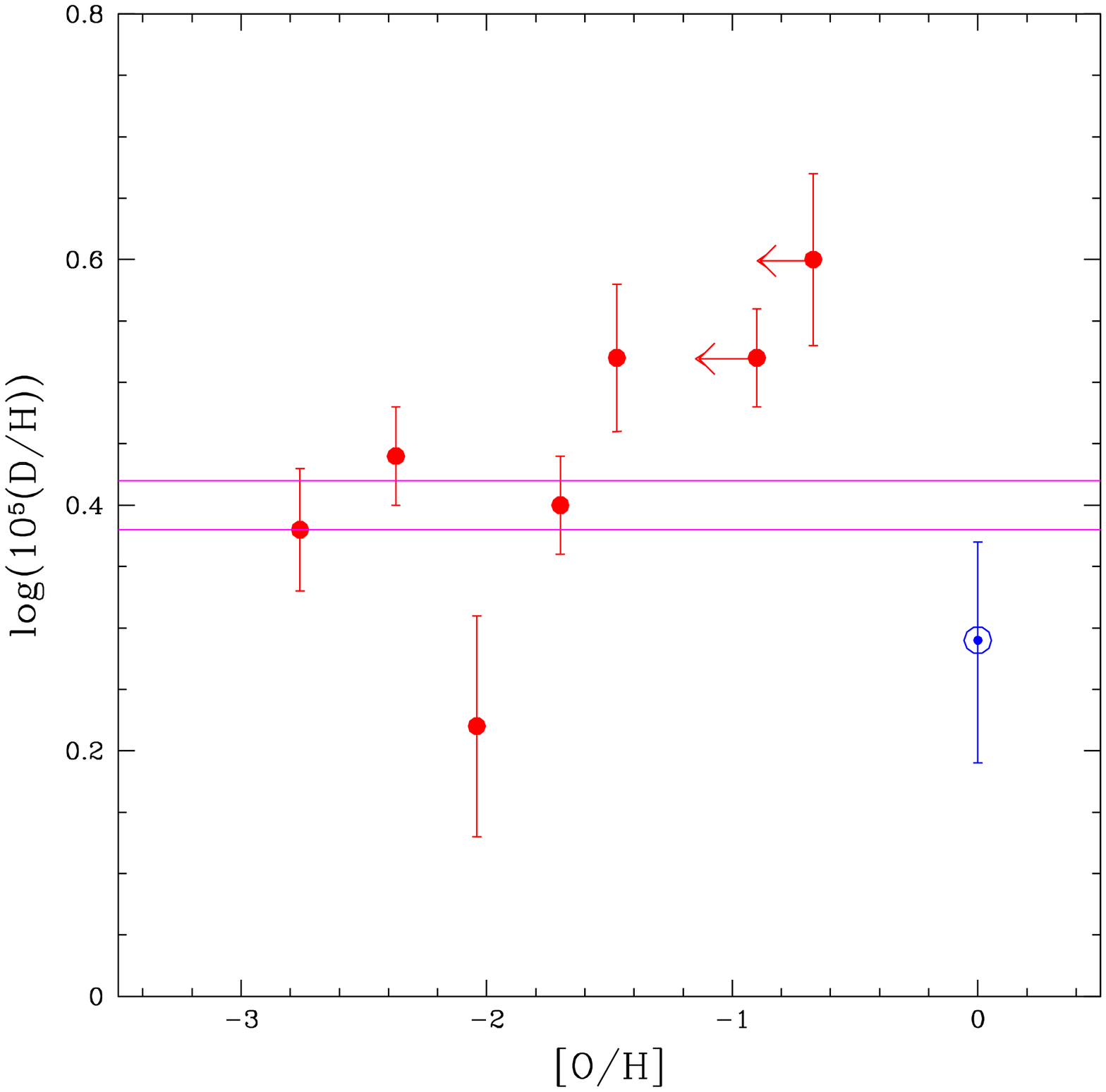} 
 \caption{The logs of the deuterium abundances, 
 $y_{\rm D} \equiv 10^{5}$(D/H), observed in high-$z$, 
 low-Z QSO Absorption Line Systems (\cite{pettini}), 
 as a function of the corresponding oxygen abundances.  
 For comparison, the solar deuterium and oxygen 
 abundances are shown (\cite{gg}).  The band 
 indicated by the solid lines is the 68\% range of the 
 SBBN-predicted primordial D abundance using 
 the CMB-determined baryon density parameter 
 (see \S 3).}
   \label{d}
\end{center}
\end{figure}

The deuterium abundance is determined by comparing the 
\hi~and \di~column densities inferred from observations 
of absorption of radiation from background UV sources 
by intervening gas.  In searching for the Deuterium 
Plateau the relelvant data is provided by observations 
of high-redshift, low-metallicity, QSO Absorption Line 
Systems (QSOALS).  Unfortunately, at present there are 
only seven, relatively reliable D abundance determinations 
\cite{pettini}, which are shown in Figure \ref{d}.  

The weighted mean of the seven D abundances is log($y_{\rm 
DP}) = 0.45$.  However, as may be seen from the Figure \ref{d}, 
only three of the seven abundances lie within $1 \sigma$ of 
the mean.  Indeed, the fit to the weighted mean of these 
seven data points has a $\chi^{2} = 18$ ($\chi^{2}/dof = 3$).  
Either the quoted errors are too small or, one (or more) 
of the determinations is wrong, perhaps contaminated by 
unidentified (and, therefore, uncorrected) systematic errors.  
In the absence of evidence identifying the reason(s) for 
such a large dispersion, the best that can be done at present 
is to adopt the mean D abundance and to inflate the error 
in the mean in an attempt to account for the unexpectedly 
large dispersion among the D abundances (\cite{steigman07}).  
\beq
{\rm log}(y_{\rm DP}) \equiv 0.45 \pm 0.03.
\label{ydp}
\eeq
For this relic D abundance, SBBN predicts 
that the baryon density parameter is
\beq
\eta_{10}({\rm SBBN}) = 5.80 \pm 0.28,
\label{etasbbn}
\eeq
corresponding to a baryon mass density $\Omega_{\rm B}h^{2} 
= 0.0212 \pm 0.0010$.

For $\eta_{10}({\rm SBBN})$, the SBBN-predicted 
abundances of the remaining light nuclides are
\beq
y_{3\rm P} \equiv 10^{5}(^{3}{\rm He/H})_{\rm P}
= 1.07 \pm 0.04,
\label{3hesbbn}
\eeq
\beq
{\rm Y}_{\rm P} = 0.2482 \pm 0.0007,
\label{4hesbbn}
\eeq
\beq
[{\rm Li}]_{\rm P} \equiv 12 + {\rm log (Li/H)}_{\rm P} 
= 2.67^{+0.06}_{-0.07}.
\label{7lisbbn}
\eeq

\begin{figure}[b]
\begin{center}
 \includegraphics[width=3.4in]{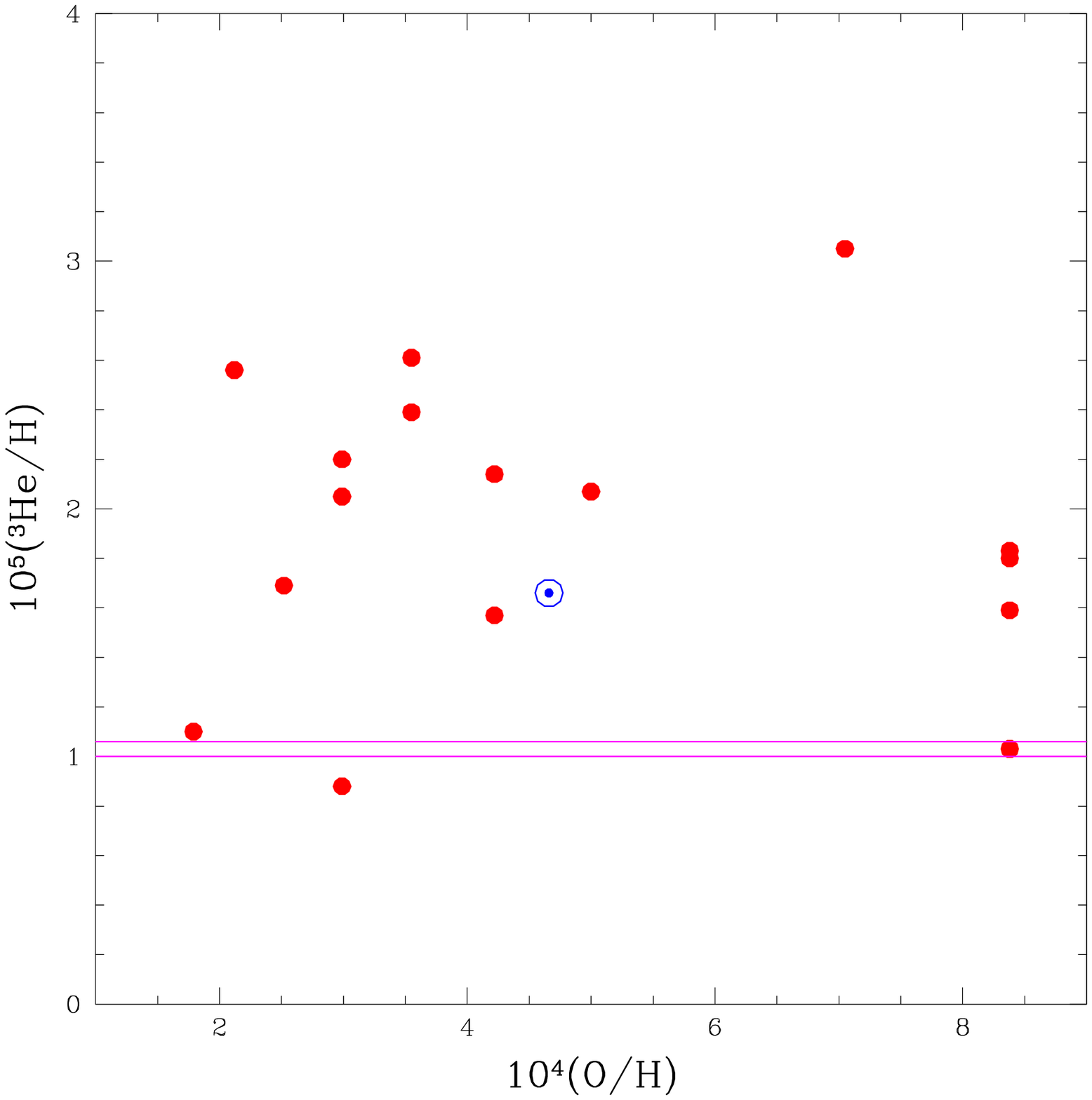} 
 \caption{The $^{3}$He abundances, $y_{3} \equiv 
 10^{5}$($^{3}$He/H), observed in Galactic \hii 
 regions from \cite{bania} are shown as a function 
 of the corresponding oxygen abundances.  For 
 comparison, the solar helium-3 and oxygen 
 abundances are shown.  The band indicated by 
 the solid lines is the 68\% range of the SBBN 
 prediction for the primordial \3he abundance 
 using the CMB-determined baryon density 
 parameter (see \S 3).}
   \label{3he}
\end{center}
\end{figure}

\subsection{Consistency Of SBBN?}

Having used the deuterium observations along with 
the predictions of SBBN to determine the baryon 
density parameter, we may now ask if the observed 
abundances of \3he, \4he, and \7li are consistent 
with their SBBN-predicted primordial values.\\ 

{\underline{\it Helium - 3}}.

The post-BBN evolution of \3he is model dependent 
and, considerably more complicated than that of D.  
Overall, the \3he abundance is expected to increase 
during Galactic chemical evolution (see, {\it e.g.}, 
\cite{rood,rst}).  Observations of \3he are limited 
to the relatively evolved \hii~regions in the Galaxy 
(\cite{bania}).  The data are shown in Figure\,\ref{3he}, 
where the observed \3he abundances are plotted versus 
the corresponding \hii~region oxygen abundances.   The 
data reveal a {\it minimum} (primordial?) \3he abundance 
which is consistent with the SBBN prediction.  While 
the higher observed abundances support the expectation 
of net post-BBN production of \3he, the absence of a 
correlation with the oxygen abundances is puzzling.\\

{\underline{\it Helium - 4}}.

\begin{figure}[b]
\begin{center}
 \includegraphics[width=3.4in]{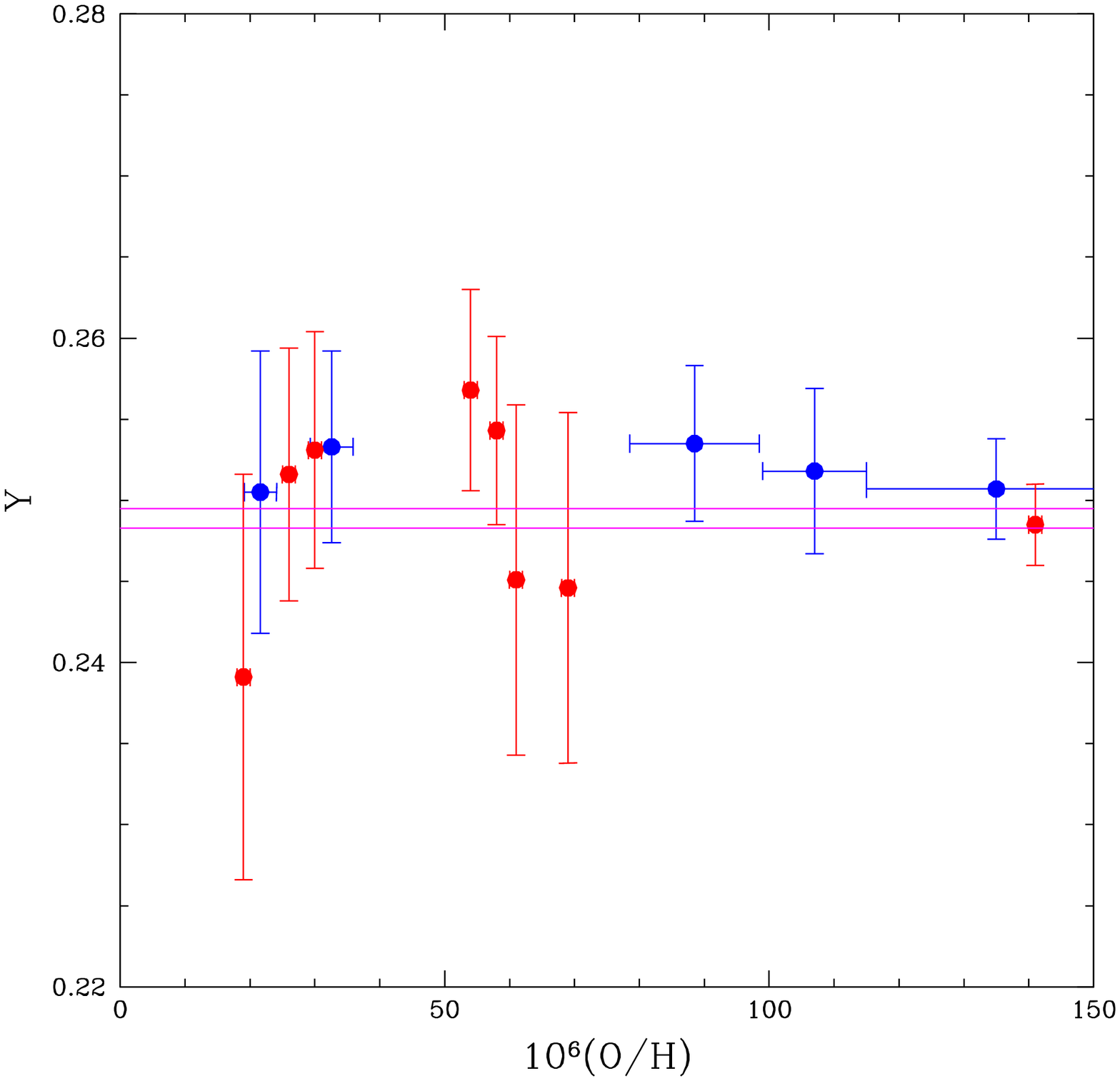} 
 \caption{The \4he mass fractions, Y, derived from 
 a selected sample (see the text) of low metallicity, 
 extragalactic \hii~regions, as a function of the 
 corresponding \hii~region oxygen abundances.  
 The blue filled circles are from \cite{plp} and the 
 red filled squares are from \cite{os}.  The band 
 indicated by the solid lines is the 68\% range of 
 the SBBN prediction for the primordial \4he mass 
 fraction using the CMB-determined baryon density 
 parameter (see \S 3).}
   \label{4he}
\end{center}
\end{figure}

The primordial abundance (mass fraction) of \4he 
is inferred from observations helium and hydrogen 
recombination lines from metal-poor, extragalactic
\hii~regions (Blue Compact Galaxies: BCDs).  In 
using these data to determine the primordial helium 
abundance, the systematic errors (underlying stellar 
absorption, temperature and density inhomogeneities, 
ionization corrections, atomic emissivities, etc.) 
dominate over the statistical errors and the uncertain 
extrapolation to zero metallicity.  In my opinion, the 
uncertainty in \Yp~is $sigma({\rm Y}_{\rm P}) \approx 
0.006$ and {\bf not} $\sigma({\rm Y}_{\rm P}) < 0.001$, 
as claimed in some published papers.  Therefore, rather 
than show the helium abundances inferred from observations 
of hundreds of BCDs, in Figure \ref{4he} are shown the 
handful of helium abundances determined from careful 
observations of a few \hii~regions where attention has 
been paid to some but, even here, not all, sources 
of systematic uncertainties (\cite{os,plp}).  The seven 
\cite{os} \hii~regions are consistent with {\bf no} 
correlation between the helium and oxygen abundances, 
leading to a weighted mean, Y$_{\rm OS} = 0.2500 \pm 
0.0030$.  The same is true for the five \hii~regions 
studied by \cite{plp}, with Y$_{\rm PLP} = 0.2517 \pm 
0.0043$, where the PLP statistical and systematic 
errors have been added linearly.  The independent 
analyses of OS and PLP agree.  The surprising absence 
of evidence for statistically significant slopes in 
their Y versus O/H relations prevents an extrapolation 
to zero metallicity in order to find Y$_{\rm P}$.  
However, the weighted means do provide {\it upper 
bounds} to Y$_{\rm P}$: Y$_{\rm P} < \langle Y \rangle$.  
As may be seen by comparison with eq.~\ref{4hesbbn},
these data are consistent with the SBBN prediction.\\

{\underline{\it Lithium - 7}}.

\begin{figure}[b]
\begin{center}
 \includegraphics[width=3.4in]{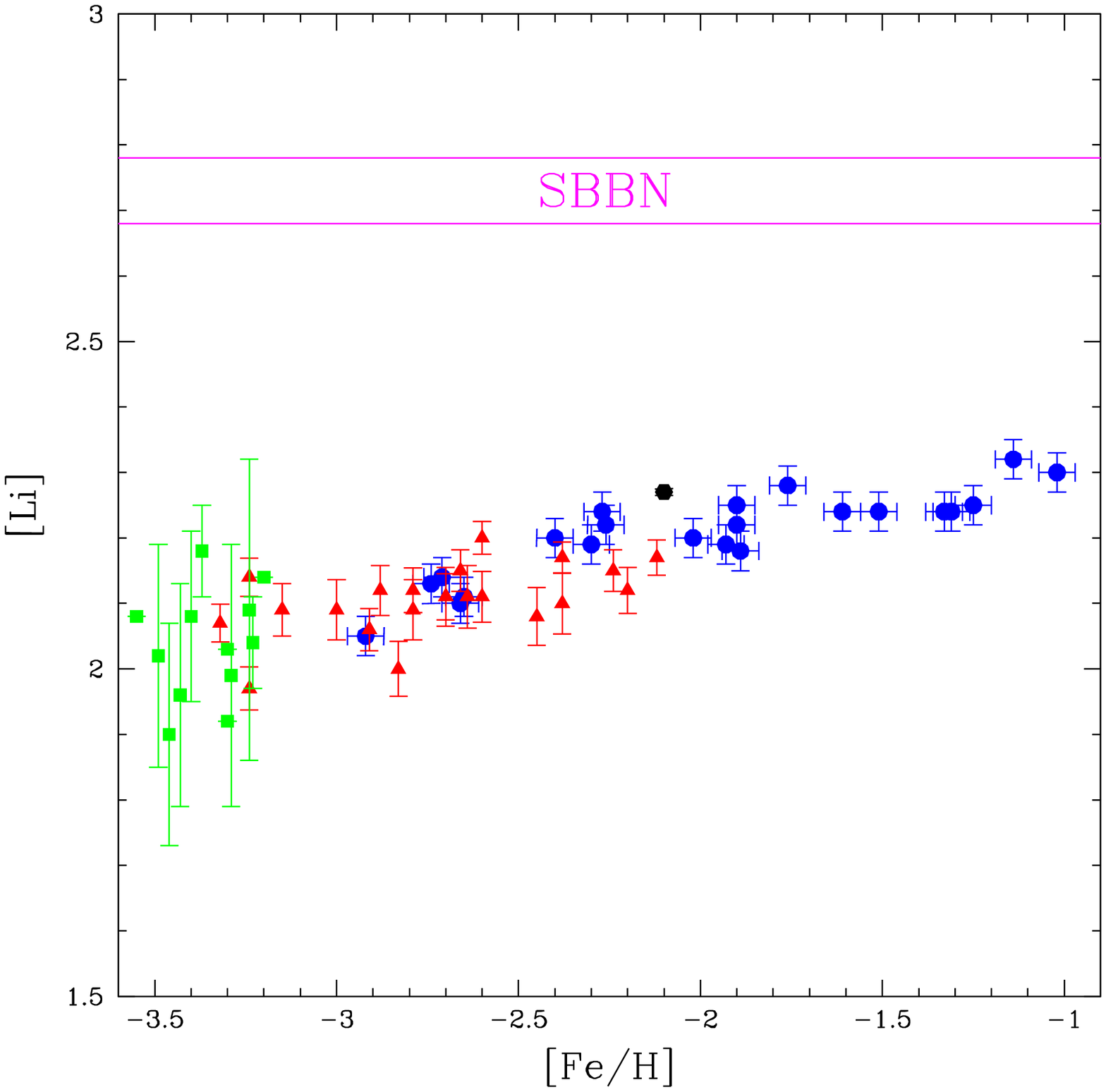} 
 \caption{The lithium abundances, [Li] $\equiv 12 
 + $log(Li/H), derived from observations of low 
 metallicity Galactic halo and globular cluster stars 
 as a function of the iron abundance (relative to 
 solar).  Blue filled circles (\cite{asplund06}), red, 
 filled triangles (\cite{boesgaard05}), green filled 
 squares (\cite{aoki09}).  The black filled circle 
 (\cite{lind09}) is for the globular cluster NGC6397.  
 The band indicated by the solid lines is the 68\% 
 range of the SBBN prediction for the primordial 
 \7li abundance using the CMB-determined baryon 
 density parameter (see \S 3).}
   \label{li}
\end{center}
\end{figure}

Like deuterium, lithium (\6li and \7li) is fragile.  
In contrast to deuterium, post-BBN lithium is produced 
via Cosmic Ray Nucleosynthesis and by some stars (see 
these Proceedings).  This is confirmed by Galactic 
observations of lithium as a function of metallicity.  
It is therefore expected that in the limit of low 
metallicity the lithium abundance should approach 
a plateau, the ``Spite Plateau".  However, while the 
primordial abundances of \3he and \4he inferred from 
the observational data are consistent with the SBBN 
predicted abundances based on the deuterium abundance, 
\7li poses a severe problem.  As may be seen from 
Figure \ref{li}, the lithium abundances derived from 
observations of the most metal-poor halo and globular 
cluster stars in the Galaxy lie well below the 
SBBN-predicted value (see eq.~\ref{7lisbbn}).   The 
discrepancy between the prediction and the observations 
is a factor of $\sim 3 - 5$.  In addition, at the 
lowest iron abundances, the lithium abundances appear 
to be decreasing with metallicity.  Where is the Spite 
Plateau?  What is the value of [Li]$_{\rm P}$?  

Thus, although the predictions and observations
of D, \3he, and \4he are consistent with SBBN,
lithium is a problem.  Setting lithium aside, 
we may ask if SBBN is consistent with the CMB?

\section{SBBN With The CMB-Inferred Baryon 
Density Parameter}

The CMB temperature anisotropy spectrum depends 
on the baryon density parameter $\eta_{\rm B}$ 
(see the contribution by Dunkley in this 
volume).  From the WMAP data \cite{dunkley09} 
find $\Omega_{\rm B}h^{2} = 0.02273 \pm 
0.00062$, which corresponds to (\cite{steigman06})
\beq
\eta_{10}({\rm CMB}) = 6.226 \pm 0.170.
\eeq
The baryon density parameters inferred from deuterium 
and SBBN, when the Universe was $\sim 20$ minutes old, 
and from the CMB, last scattered some 400 thousand 
years later, agree within $\sim 1.5\sigma$ (the glass 
is half full).  SBBN and the CMB are consistent 
(modulo the lithium problem).

It is interesting to check the consistency of SBBN and 
the CMB by comparing the SBBN-predicted primordial light 
nuclide abundances determined using the CMB value of 
baryon density parameter to the observations.  These
comparisons are shown by the horizontal bands in Figures 
\ref{d}-\ref{li}.  For SBBN with $\eta_{\rm B}({\rm CMB})$,
\beq
{\rm log}(y_{\rm DP}) = 0.40 \pm 0.02 ~~(y_{\rm DP} = 
2.52 \pm 0.13),
\eeq
\beq
y_{3{\rm P}} = 1.03 \pm 0.03,
\eeq
\beq
{\rm Y}_{\rm P} = 0.2489 \pm 0.0006,
\eeq
\beq
[{\rm Li}]_{\rm P} = 2.74 \pm 0.05.
\eeq
The SBBN/CMB-predicted abundances of D, \3he, and 
\4he are consistent with their observationally-inferred 
primordial values, but the lithium discrepancy is 
exacerbated.  SBBN (\nnu~= 3) and the CMB are consistent 
(but lithium is a problem!).

\section{Non-Standard Big Bang Nucleosynthesis 
(nSBBN): \nnu~$\neq 3$}
\label{nsbbn}

For non-standard BBN (nSBBN) with \nnu~$\neq 3$, the relic 
abundances of the light nuclides are functions of two 
parameters, $\eta_{\rm B}$ and \nnu.  First, consider 
deuterium.  The nSBBN primordial abundance is predicted 
to vary as $y_{\rm DP} \propto \eta_{\rm D}^{-1.6}$, where 
$\eta_{\rm D} = \eta_{\rm D} (\eta_{10},$N$_{\nu})$ (see, 
{\it e.g.}, \cite{ks04,steigman07}).  It is interesting to 
explore the consequences of using the CMB (\cite{dunkley09}) 
to fix $\eta_{10}$ and the observed primordial D abundance 
to determine $\eta_{\rm D}$.  This leads to a combined BBN 
and CMB prediction for \nnu.  For log$(y_{\rm D}) = 0.45 
\pm 0.03$ and $\eta_{10}({\rm CMB}) = 6.23 \pm 0.17$, 
\nnu~$= 4.0 \pm 0.7$.  Here, the relative insensitivity 
of $y_{\rm DP}$ to \nnu~has amplified the small difference 
between $\eta_{10}$ and $\eta_{\rm D}$ into a relatively 
large value (and uncertainty) of \Deln~$= 1.0 \pm 0.7$.  
Although the central value of the effective number of 
neutrinos determined this way is \nnu~$\neq 3$, this 
result is consistent with \nnu~= 3 at $\sim 1.4\sigma$.  
Using this value of \nnu~along with $\eta_{10}({\rm CMB})$, 
how do the predicted BBN abundances of the remaining 
light nuclides compare with their observationally 
inferred primordial values?  Here, I concentrate on 
the two key elements, \4he and \7li (by construction, 
D is de facto consistent).

For this combination of $\eta_{10}$ and \nnu, the primordial
\4he mass fraction is \Yp~$= 0.261 \pm 0.009$.  Here, the 
sensitivity of \Yp~to \nnu~has amplified the small difference
between $\eta_{10}$ and $\eta_{\rm D}$ into a relatively 
large value (and uncertainty) of \Yp.  As may be seen 
from Figure~\ref{4he}, within the large uncertainty of 
this prediction, the very high central value is 
consistent with the data. 

The BBN-predicted abundances of D and \7li are very 
tightly correlated, both for \nnu~= 3 and \nnu~$\neq 3$ 
(\cite{ks04,steigman07}).  As a result, even for this 
example of nSBBN, the predicted primordial lithium 
abundance is very similar to its SBBN value, [Li]$
_{\rm P} = 2.70^{+0.05}_{-0.06}$, in conflict with 
the observational data in Figure~\ref{li}.

Thus, although this variant of nSBBN is consistent 
with D, \3he, and \4he, the lithium problem persists!  
Nonetheless, this example illustrates the potential 
value of combining BBN and the CMB to constrain and 
test non-standard models of particle physics and 
cosmology.  A slightly different variant of this 
approach is presented in the next section.

\section{Using \4he And The CMB To Constrain \nnu}

Of the light nuclides synthesized during BBN, the \4he 
mass fraction is most sensitive to non-standard physics 
(\nnu~$\neq 3$).  Indeed, for $|$\Deln$| \la 1$, $\Delta$Y
$_{\rm P} \approx 0.013$\Deln, so that a good bound (small 
uncertainty) to \Yp~would result in a tight constraint 
on \nnu.  According to \cite{ks04} and \cite{steigman07}, 
using $\eta_{10}$(CMB)$ = 6.22 \pm 0.16$, a very good 
fit to \Yp, is 
\beq
{\rm Y}_{\rm P} \approx 0.2486 \pm 0.0007 + 
0.013\Delta{\rm N}_{\nu}.
\eeq
As mentioned earlier, the present uncertainty in the 
observationally inferred value of \Yp~is dominated by 
systematic errors.  To illustrate the potential value 
of an accurate determination of \Yp, let's adopt the 
weighted mean of the \cite{os} helium abundances as an 
{\it upper bound} to the primordial \4he mass fraction: 
\Yp~$< \langle Y \rangle_{\rm OS} = 0.2500 \pm 0.0030$.   
Comparing this BBN prediction of \Yp~with the upper 
bound inferred from the data leads to an {\it upper 
bound} on the effective number of neutrinos,
\beq
\Delta{\rm N}_{\nu} < 0.11 \pm 0.24~~({\rm N}_{\nu} 
< 3.11 \pm 0.24).
\eeq
If, instead, we had adopted the weighted mean of the 
\cite{plp} helium abundances, we would have found 
$\Delta$N$_{\nu} < 0.24 \pm 0.33$ (\nnu~$= 3.24 \pm 0.33$).  
Just a few good \hii~region \4he abundances are all that 
is needed to obtain a very strong constraint on \nnu.  
For these combinations of $\eta_{10}$(CMB) and \nnu, the 
bounds to the D and \3he abundances are consistent with 
the data, while lithium remains a problem.
 
\section{Challenges}

SBBN (\nnu~= 3) is consistent with the CMB and with the 
observationally-inferred primordial abundances of D, \3he, 
and \4he, but \7li poses a problem.  SBBN and the CMB in 
combination allow, but also constrain, some models of 
non-standard physics.  The challenges facing BBN, largely 
observational, makes the timing of this meeting ideal.  
Having had the luxury of being one of the first speakers, 
I will end by presenting my list of challenges to those 
who follow.\\
{1. Why is the spread in the observed deuterium abundances
so large?}\\
{2.  Why are the observed \3he abundances uncorrelated with
either the oxygen abundances or with the distance from the
center of the Galaxy?}\\
{3. What are the sources (and the magnitudes) of the systematic
errors in \Yp~and, are there observing strategies to reduce 
them?}\\
{4. What is the primordial abundance of \7li (and of \6li)?}


\begin{thebibliography}{}

\bibitem[Aoki \etal\ (2009)]{aoki09}
{Aoki, W. \etal} (2009),
\textit{ApJ}, In Press, (arXiv:0904.1448).

\bibitem[Asplund \etal\ (2006)]{asplund06}
{Asplund, M., \etal} 2006,
\textit{ApJ}, 644, 229.

\bibitem[Bania, Rood, \& Balser (2002)]{bania}
{Bania, T.M., Rood, R.T., \& Balser, D.S.} 2002,
\textit{Nature}, 415, 54.

\bibitem[Boesgaard \etal\ (2005)]{boesgaard05}
{Boesgaard, A.M., Stephens, A., \& Deliyannis, C.P.} 2005,
\textit{ApJ}, 633, 398. 

\bibitem[Dunkley \etal\ (2009)]{dunkley09}
{Dunkley, J., \etal} 2009,
\textit{ ApJS}, 180, 306.

\bibitem[Epstein, Lattimer, \& Schramm (1976)]{els}
{Epstein, R. I., Lattimer, J. M., \& Schramm, D. N.} 1976,
\textit{Nature}, 265, 219.

\bibitem[Geiss \& Gloeckler (1998)]{gg}
{Geiss, J. \& Gloeckler, J.G.} 1998,
\textit{Space Sci. Rev.} 84, 239.

\bibitem[Kneller \& Steigman (2004)]{ks04}
{Kneller, J.P. \& Steigman, G.} 2004,
\textit{New J. Phys.}, 6, 117.

\bibitem[Lind \etal\ (2009)]{lind09}
{Lind, K., \etal} 2009,
\textit{A\&A},503, 545.

\bibitem[Olive \& Skillman (2004)]{os}
{Olive, K. A. \& Skillman, E. D.} 2004,
\textit{ApJ}, 617, 29

\bibitem[Peimbert, Luridiana, \& Peimbert (2007)]{plp}
{Peimbert, M., Luridiana, V. \& Peimbert, A.} 2007,
\textit{ApJ}, 666, 634

\bibitem[Pettini \etal\ (2008)]{pettini}
{Pettini, M., Zych, B.J., Murphy, M.T., 
Lewis, A., \& Steidel, C.C.} 2008,
\textit{MNRAS}, 391, 1499.

\bibitem[Reeves \etal\(1973)]{deut}
{Reeves, H., Audouze, J., Fowler, W. A., \& Schramm,
D. N.} 1973,
\textit{ApJ}, 179, 909.

\bibitem[Rood (1972)]{rood}
{Rood, R.T.} 1972,
\textit{ApJ}, 177, 681.

\bibitem[Rood, Steigman, \& Tinsley (1976)]{rst}
{Rood, R.T., Steigman, G., \& Tinsley, B.M.} 1976,
\textit{ApJL}, 207, L57.

\bibitem[Steigman (2006)]{steigman06}
{Steigman, G.} 2006,
\textit{JCAP}, 10, 016.

\bibitem[Steigman (2007)]{steigman07}
{Steigman, G.} 2007,
\textit{Ann. Rev. Nucl. Part. Sci.}, 57, 463.

\end{thebibliography}
\end{document}